\documentclass[twocolumn,amsmath,amssymb,showpacs,prl,aps,longbibliography]{revtex4-1}
\pdfoutput=1
\usepackage[bookmarks=false,linkcolor=blue,urlcolor=blue,colorlinks,citecolor=blue]{hyperref}
\usepackage{graphicx}
\usepackage{bm}
\usepackage{amssymb}
\usepackage{amsmath}
\usepackage{amsfonts}
\usepackage{epsfig}
\usepackage{stackrel}
\usepackage{paralist}
\usepackage[rgb]{xcolor}
\usepackage{todonotes}
\usepackage{mathrsfs}
\usepackage{tabularx}

\newcommand {\myvec}[1] {{\mbox{\boldmath $#1$}}}

\def\be{\begin{equation}}
\def\ee{\end{equation}}
\def\ba{\begin{align}}
\def\ea{\end{align}}
\newcommand{\bmat}{\begin{bmatrix}}
\newcommand{\emat}{\end{bmatrix}}

\begin{document}
\title{Topological transitions and fractional charges induced by strain and magnetic field in carbon nanotubes}
\author{Yonathan Efroni}
\affiliation{Department of Condensed Matter Physics, Weizmann Institute of Science, Rehovot 7610001, Israel}
\author{Shahal Ilani}
\affiliation{Department of Condensed Matter Physics, Weizmann Institute of Science, Rehovot 7610001, Israel}
\author{Erez Berg}
\affiliation{Department of Condensed Matter Physics, Weizmann Institute of Science, Rehovot 7610001, Israel}
\affiliation{Department of Physics, James Frank Institute, University of Chicago, Chicago, Illinois 60637, USA}

\begin{abstract}
We show that carbon nanotubes (CNT) can be driven through a topological phase transition using either strain or a magnetic field. This can naturally lead to Jackiw-Rebbi soliton states carrying fractionalized charges, similar to those found in a domain wall in the Su-Schrieffer-Heeger model, in a setup with a spatially inhomogeneous strain and an axial field. Two types of fractionalized states can be formed at the interface between regions with different strain: a spin-charge separated state with integer charge
 and spin zero (or zero charge and spin $\pm \hbar/2$), and a state with charge $\pm e/2$ and spin $\pm \hbar/4$. The latter state requires spin-orbit coupling in the CNT. We show that in our setup, the precise quantization of the fractionalized interface charges is a consequence of the symmetry of the CNT under a combination of a spatial rotation by $\pi$ and time reversal. 
\end{abstract}
\maketitle

\emph{Introduction.--}
Charge fractionalization is one of the most fascinating manifestation of emergent behavior in condensed matter physics. This phenomenon results from a subtle interplay of quantum many-body physics and topology. Although fractional charges were theoretically predicted to arise in different settings, only a handful of confirmed experimental realizations exist, the most prominent being fractional charges in quantum Hall states~\cite{Tsui1982,Laughlin1983,de-Picciotto1997,Saminadayar1997}.

The earliest theoretical prediction of charge fractionalization was given by Jackiw and Rebbi~\cite{JackiwRebbi} in the context of relativistic field theory. They showed that a one-dimensional (1D) Dirac equation with a spatially varying mass has a zero-energy eigenstate whenever the mass changes sign. This is a ``half fermion'' state, which carries a fermion charge of $e/2$ relative to the background filled Dirac sea. The solid state equivalent was suggested by Su, Schrieffer and Heeger (SSH)  \cite{SSHModel,SSHModelExcitations} who, motivated by the structure of polyacetylene, considered a 1D dimerized chain of electrons  (see also Refs.~\cite{rice1979charged, Brazovskii1981}). In their model, a Jakiw-Rebbi soliton carrying sharply defined fractionalized quantum numbers~\cite{Kivelson1982} emerges at a domain wall between the two dimerization states. The Jakiw-Rebbi soliton state plays also an important role in the theory of topological insulators ~\cite{Hasan2010,Qi2011}. Jakiw-Rebbi zero modes, and the geometric Zak phase~\cite{ZakPhase} of the underlying Bloch wavefunctions, were observed in photonic and cold atomic systems~\cite{ObservationOfToplogicalSolitonState,ObservationOfToplogicalPhotonicWalks,ZakPhaseMeasurement}. However, a direct observation of a Jackiw-Rebbi soliton state and its
associated fractional charge in solid state systems is still lacking.

\begin{figure}[h]
\centering
\includegraphics[width=1.05\columnwidth]{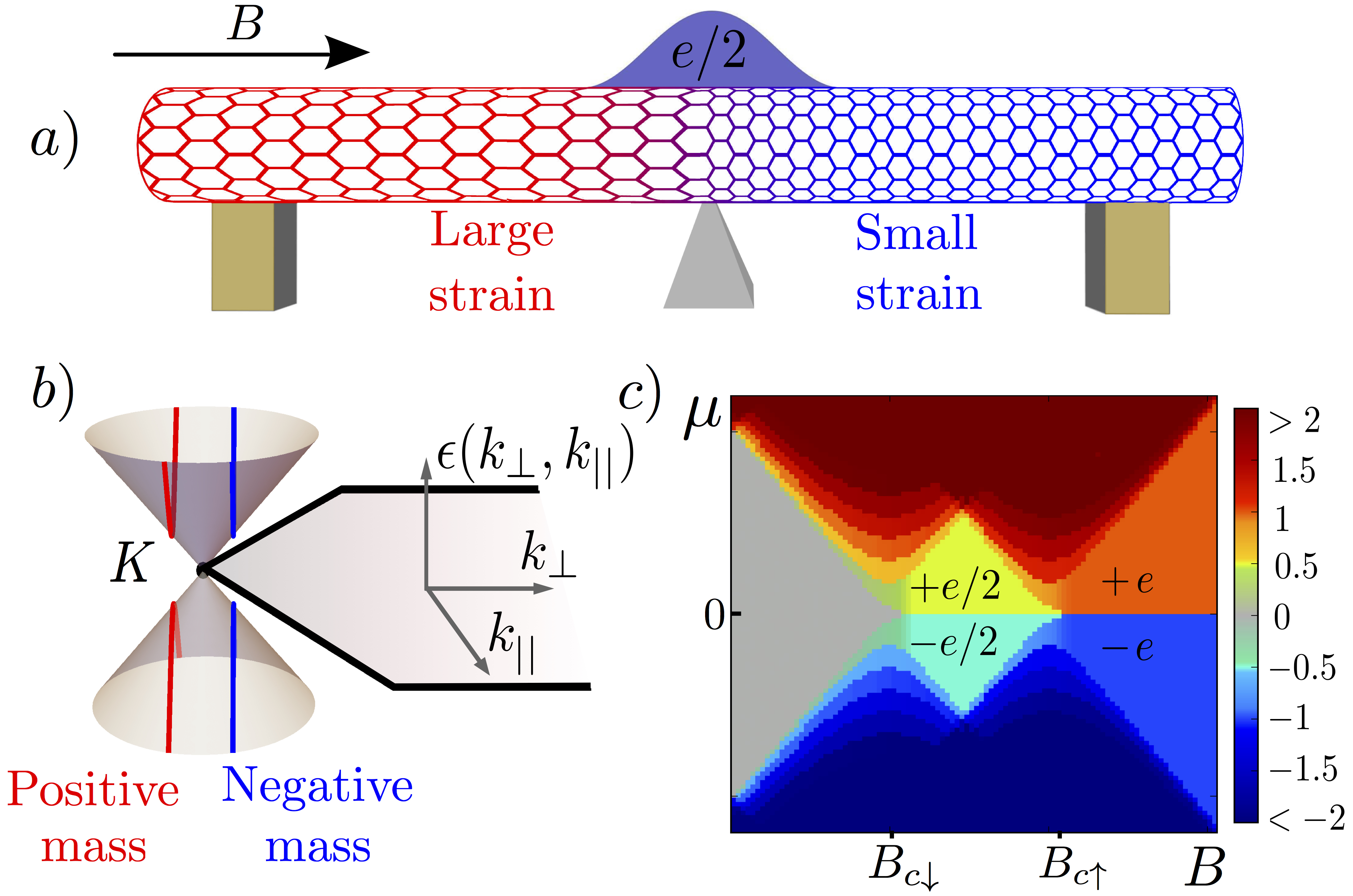}
\caption{ a) The proposed system. A suspended CNT is placed on a wedge-shaped pillar, causing the strain of the CNT at the left side(red) to be different then at the right side(blue). An axial magnetic field is applied. At a certain range of field, a Jackiw-Rebbi soliton state with a fractionalized charge is formed in the middle region.
b) The energy dispersion of the CNT can be understood as a 1D cut through the dispersion of the honeycomb lattice. The position of the cut depends on strain and magnetic field. Thus, as the field is varied, one of the CNT sub-bands may cross through the Dirac point, changing the sign of its mass term. c) Predicted charge in the middle third of the CNT, as a function of magnetic field, $B$, and chemical potential, $\mu$. (See text for details of the simulation.) In a range of fields and chemical potentials, the charge is $\pm e/2$; at higher fields, a spin-charge separated state with charge $\pm e$ and spin zero is formed.
}
\label{fig:suggested_system}
\centering
\end{figure}


In this work, we propose a simple, robust experiment for realizing Jackiw-Rebbi fractionally charged states in carbon nanotubes (CNT).
The suggested experimental setup (Fig.~\ref{fig:suggested_system}a)
consists of a metallic CNT suspended between two contacts and touching a wedge-shaped pillar near its middle.
In a metallic CNT the quantization of the perpendicular momentum is such that the line of allowed momenta cuts through the Dirac point, leading to a 1D Dirac electronic dispersion~\cite{ilani2010electron}.
However, real CNTs always have intrinsic tension that shifts the quantization condition away from the Dirac point, opening a small gap ($\sim1-100$meV)~\cite{minot2004determination, SO_CNT_3}. By applying a magnetic field parallel to the CNT axis it is possible to shift the quantization condition such that the gap closes and reopens after crossing the Dirac point~\cite{SO_CNT_meas_3,deshpande2009mott,Meerwaldt2012}. Since the strain on both sides of the wedge is not necessarily equal, the gap in these two regions generically closes at different fields. Thus, there is a range of magnetic fields where the Dirac masses in the two regions \emph{have an opposite sign} (Fig.~\ref{fig:suggested_system}b).  We will show that in this range, a {robust} SSH-like soliton state carrying a fractionalized charge emerges at the interface.

Fig.~\ref{fig:suggested_system}c shows the predicted charge stability diagram for the system, 
that carries fingerprints of the fractionally charged soliton states. The colormap represents the charge integrated over a spatial region around the wedge, as a function of the axial magnetic field, $B$, and chemical potential, $\mu$.
For $\mu$ near the middle of the semiconducting gap of the CNT, three distinct regions of $B$ are shown: $B<B_{c\downarrow}$, where the charge is zero ($B_{c\downarrow}$ is the field of the topological transition for spin $\downarrow$ electrons, see Eq.~\eqref{eq:Bc} below); $B_{c\downarrow}<B<B_{c\uparrow}$, where the charge is $\pm e/2$; and $B_{c\uparrow}<B$, where the system exhibits ``spin-charge separation:'' the charge in this region is $\pm e$, while the spin of the soliton is zero. $B_{c\downarrow}$ and $B_{c\uparrow}$ are different due to spin-orbit coupling in the CNT; this effect is crucial for realizing the $\pm e/2$ state.

As we will show below, the precise quantization of the charge localized in the middle region of the CNT
in units of $e/2$ is a consequence of the symmetry of the CNT under a rotation by $\pi$ followed by time reversal. This symmetry is present in both chiral and non-chiral CNTs.

For a magnetic field in the range $B_{c\downarrow}<B<B_{c\uparrow}$, a precisely quantized fractional charge appears also \emph{at the edge} of the CNT (see~\cite{SOM}). In-gap edge states may appear, as well~\cite{EdgeStatesWithDiracEquation,PhysRevB.93.195442}. Our setup has the practical advantage that the fractional charge is realized near the middle of the CNT; the edges tend to be less clean and well-controlled compared to the bulk. Moreover, in our setup the soliton is far from any metallic contacts that can bend the CNT's bands due to the difference in work function, masking the fractional charge.

\emph{Carbon nanotunbe model.--}
The low-energy effective Hamiltonian of a CNT is given by
 $h(\myvec{k})=\hbar v_F(\xi \sigma_x k_x+\sigma_y k_y)$.
Here, $v_F$ is the velocity of the Dirac point, $\sigma_x, \sigma_y$ are Pauli matrices acting in the sublattice (A-B) space, and $\xi = +1 (-1)$ corresponds to the  $K'$ ($K$) valley, respectively. 

The CNT is specified by the chiral vector,  $\myvec{c}=n_1\myvec{a_1}+n_2\myvec{a_2}$, denoted by $c=(n_1,n_2)$, which connects two carbon atoms of the parent graphene sheet~\cite{BookCNT}. The perpendicular momentum to the tube is quantized due to its finite diameter.
In an ideal metallic CNT ($n_1-n_2 \in 3\mathbb{Z} $), the lines of allowed momenta cross the $K$, $K'$ points. 
An axial magnetic field $B$ and curvature effects in CNTs shift the allowed momenta lines away from $K$, $K'$~\cite{minot2004determination,TuningCNTGapStrain}. The low-energy Hamiltonian takes the form~\cite{SO_CNT_3}:
 \begin{eqnarray}
 h(k_{\parallel})&=&\hbar v_F(\xi \sigma_x \frac{2 \pi \phi}{ |\myvec{c}| \phi_0} +\sigma_y k_{\parallel}) \nonumber \\
 &+& \delta t\sigma_x - (\Delta^{SO}_{\mathrm{o}}\sigma_x + \Delta^{SO}_{\mathrm{z}}) \xi S_z.
 \label{eq:3}
 \end{eqnarray}
Here, $\phi = B a^2 (|\mathbf{c}|/2\pi)^2$ is the flux through the cross section of the CNT (where $a$ is the lattice spacing), $\phi_0 = h/e$ is the flux quantum, $k_\parallel$ is the momentum parallel to the CNT, $S_z$ is the spin along the CNT axis, $\Delta^{SO}_{\mathrm{o}}, \Delta^{SO}_{\mathrm{z}}$ are the strengths of the orbital and Zeeman-type spin-orbit couplings, respectively~\cite{SO_CNT_3}, and the $\delta t$ term is due to strain along the axis of the CNT. The $\Delta^{SO}_{\mathrm{o}}$, $\Delta^{SO}_{\mathrm{z}}$, and $\delta t$ terms are induced by curvature~\cite{SO_CNT_4,SO_CNT_1,SO_CNT_2,SO_CNT_5,SO_CNT_3}. $\delta t$ can be modified by applying external strain to the CNT~\cite{TuningCNTGapStrain}. We have neglected Zeeman coupling, which is smaller than the terms in Eq.~(\ref{eq:3}).



\begin{figure}[t]
	\centering
	\includegraphics[width=0.9\columnwidth]{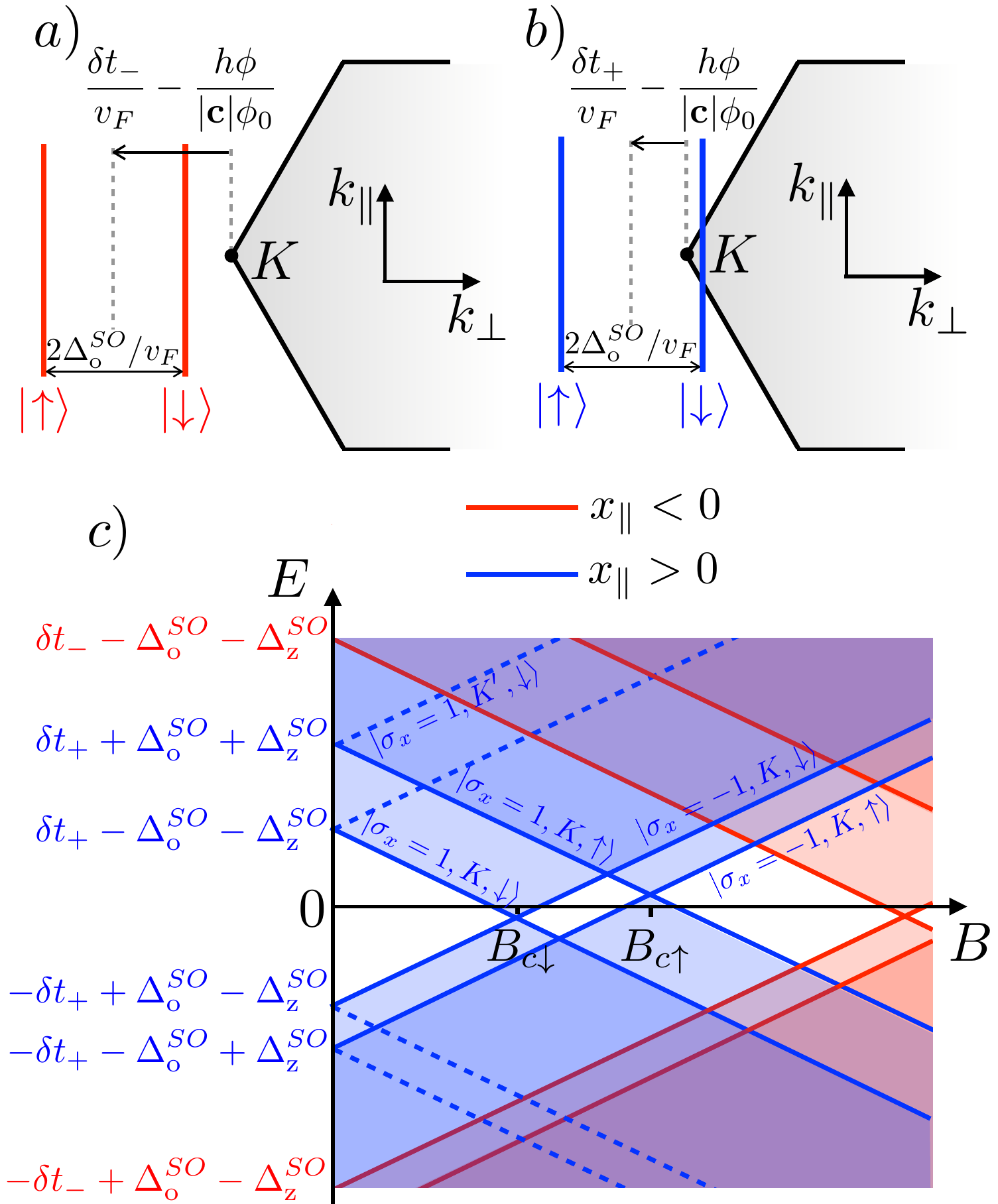}
\caption{ a,b) 
Quantization lines of the perpendicular momentum around the $K$ point for a zigzag CNT, shown in the honeycomb lattice Brillouin zone. The allowed values of $k_\perp$ depend on the strain $\delta t$ on the flux $\phi$, and on the spin [see Eq.~(\ref{eq:3})]. The blue (red) lines correspond to $x_{\parallel}<0$ ($x_\parallel>0$), respectively. One of the quantized momenta can sweep through the Dirac point in one spatial region of the system (panel a) but not in the other (panel b). This results in a Jackiw-Rebbi soliton state and a fractional charge at the interface. c) Evolution of the spectrum as a function of axial magnetic field. White regions correspond to a gap, while filled regions represent the bulk spectrum. Solid (dashed) lines correspond to the gaps of states at the $K$ ($K'$) valley. The gap closes, and the Dirac mass changes sign, at $B=B_{c\uparrow,\downarrow}$ for the $K$ valley bands with spin up and down, respectively. }
 	\label{fig:SplitDirac}
\end{figure}




\emph{Soliton states and fractional charges.--}We now consider the system shown in Fig.~\ref{fig:suggested_system}a. Here, we focus on the case of a zigzag CNT with chiral vector $(N,0)$. Similar considerations hold for other types of CNTs (see below). The tensile strain of the CNT is spatially inhomogeneous, and hence the $\delta t$ term in Eq.~(\ref{eq:3}) is $x_{\parallel}$ dependent (where $x_{\parallel}$ is the coordinate along the CNT). The wedge is at $x_{\parallel} = 0$. Across the wedge, $\delta t$ varies; we assume for simplicity that $\delta t(x_\parallel)$ is piecewise constant, and denote the values of $\delta t$ at $x_\parallel < 0$ ($x_\parallel > 0$)
by $\delta t_-$ ($\delta t_+$), respectively.


Under these assumptions, the low-energy effective Hamiltonian is written as
 \begin{eqnarray}
 H&=&\int dx_{\parallel}\,
 \Bigg[ m(\xi, S_z, x_{\parallel}) \sigma_x+
 \hbar v_F \sigma_y \left(-i\frac{\partial}{\partial x_{\parallel}} \right) \Bigg], \nonumber \\
  &+& \Delta^{SO}_z \xi S_z,
  \label{eq:JR}
 \end{eqnarray}
where $m(\xi, S_z, x_{\parallel}) = \hbar v_F \frac{2\pi \phi}{|\myvec{c}| \phi_0} \xi -  {\Delta^{SO}_{\mathrm{o}}} \xi S_z  + {\delta t(x_{\parallel})}$. Eq.~(\ref{eq:JR}) is equivalent to the Jackiw-Rebbi problem~\cite{JackiwRebbi}. For $\phi = 0$, $m$ is generically non-zero for both valleys, $\xi = \pm 1$, and in either side of the wedge. As the magnetic field increases, there is a sequence of topological phase transitions where $m$ goes through zero. For a certain range of magnetic field, the masses in the two spatial region have an opposite sign for either one or both spin flavors (See Fig.~\ref{fig:SplitDirac}a,b).
Whenever the mass of one of the spin/valley bands has an opposite sign in the two regions, there is a localized state at the interface with an associated charge of $\pm \frac{e}{2}$~\cite{JackiwRebbi}.

Fig.~\ref{fig:SplitDirac}c shows the evolution of the spectrum as a function of magnetic field. Here, we have assumed that $\delta t_- > \delta t_+ > \Delta^{SO}_{\mathrm{o}} > \Delta^{SO}_{\mathrm{z}}>0$. Then, the first two gap closing points occur for the spin up and down bands of valley $K$ at $x_\parallel > 0$.
The corresponding critical magnetic fields for spin $S_z = \uparrow,\downarrow$ are given by
\begin{equation}
B_{c,S_z} = \frac{2\pi \phi_0}{|\mathbf{c}| \hbar v_F}( \delta t_+ + \Delta^{SO}_{\mathrm{o}}S_z).
\label{eq:Bc}
\end{equation}
For $B_{c\downarrow}<B<B_{c\uparrow}$, there is a single Jackiw-Rebbi soliton state localized around the interface. The localization length of this state is $\ell \sim v_F/m(x_\parallel)$, where $m(x_\parallel)$ is the mass corresponding to valley $K$ and spin $\downarrow$ at either $x_\parallel>0$ or $x_\parallel<0$. Note that there is a different decay length to the left and to the right of the wedge. If the chemical potential is in the bulk gap, there is a well-defined charge of $\pm e/2$ distributed around the interface.
The existence of a gap in the spectrum for $B_{c\downarrow}<B<B_{c\uparrow}$ requires that $\Delta^{SO}_{\mathrm{o}} > \Delta^{SO}_{\mathrm{z}}$. The ratio of $\Delta^{SO}_{\mathrm{o}}$ and $\Delta^{SO}_\mathrm{z}$ depends on the chiral vector of the CNT (see below).

For $B>B_{c\uparrow}$, but still below the field in which the first topological phase transition occurs at $x_\parallel >0$,
there are \emph{two} soliton states bound to the interface, one for each spin. 
The charge at the interface can be either $-e$, $0$, or $e$, depending on the chemical potential. Just as in the SSH chain~\cite{RevSolitonsInPolymers}, the state of the interface exhibits ``spin-charge separation:" the charge $\pm e$ state has spin zero, while the zero charge state has spin $\pm\frac{\hbar}{2}$. In the latter state, the spin degeneracy is lifted by the $\Delta^{SO}_{\mathrm{z}}$ term.

The equivalence of our system to the SSH model can be understood at the microscopic level; see~\cite{SOM}.

\emph{Fractional charge from geometric phases.--}The presence of a fractional charge at the domain wall can be understood as a consequence of a topological invariant: the Zak phase~\cite{ZakPhase}. The Zak phase is related to the \emph{charge polarization density}.
Upon changing $\delta t$ from $\delta t_-$ to $\delta t_+$, the total change in the polarization, $\Delta P$,
is~\cite{TheoryPolarizCryst}:
\begin{equation}
\Delta P = \frac{e}{2\pi} \sum_{n \in \mathrm{occ.}}  \int^{\delta t_+}_{\delta t_-} d(\delta t) \int dk_{\parallel} \, \mathrm{Im}\langle \partial_{k_{\parallel}} u^{(n)}_{k_{\parallel},\delta t} \vert \partial_{\delta t} u^{(n)}_{k_{\parallel},\delta t} \rangle.
\label{eq:DP}
\end{equation}
Here, the $k_\parallel$ integral is over the first Brillouin zone (BZ), $\vert u^{(n)}_{k_{\parallel},\delta t} \rangle$ is the Bloch wavefunction in band $n$, and the summation is over the occupied bands. 

We can replace Eq.~(\ref{eq:DP}) by a 2D integral over the Berry curvature, $\mathcal{F}(\mathbf{k})$ in the BZ of the honeycomb lattice, with $(k_x,k_y)$ replaced with ($k_\parallel$, $k_\perp$), where  {$k_\perp = \frac{\delta t}{\hbar v_F} - \xi  \frac{2 \pi \phi}{ |\myvec{c}| \phi_0}$} [see Eq.~(\ref{eq:3})].
$\mathcal{F}(\mathbf{k})$ is non-zero only at the Dirac points: 
$\mathcal{F}(\myvec{k}) = \pi \delta(\myvec{k}-\myvec{K}) - \pi \delta(\myvec{k}-\myvec{K'}).$
Therefore,
$\Delta P$ is non-zero if during the change of $\delta t$, the dispersion of the CNT crosses one or more Dirac points. In this case, the change in polarization density is $e/2$ per Dirac point crossed. The change in the bound charge at the interface is $\Delta q = \int_{x_1}^{x_2} dx \frac{\partial \Delta P}{\partial x} = \Delta P (x_2) - \Delta P (x_1)$. Hence, it is also quantized in units of $e/2$.   


The precise quantization of the polarization us due to a spatial symmetry of the CNT. A CNT with an axial magnetic field is invariant under a rotation by $180^{\circ}$ around an axis perpendicular to the CNT axis, followed by a time reversal operation~\cite{SOM}. We denote this symmetry operation by $\mathcal{R}_\pi$. Under $\mathcal{R}_\pi$, $P\rightarrow-P$; however, since $P$ in a crystal is only defined modulo $e$, $P\mod\ e=-P\mod\ e$. Hence, the possible distinct values for $P$ are either $0$ or $e/2$, and the charge at the interface is quantized in units of $e/2$. If the interface charge is ${e}/{2}$, an extra charge of $e/2$ appears at the end of the CNT. 

Note that the system is symmetric under $\mathcal{R}_\pi$ for \emph{any} chiral vector. 
In armchair CNTs, however, both Dirac points are always crossed together, and therefore the charge at the interface is an integer multiple of $e$.

In the above discussion, we have assumed that there is no term in the Hamiltonian that breaks the symmetry between the $A$ and $B$ sublattices [a $\sigma_z$ term in Eq.~(\ref{eq:3})]. If such a term exists, then the $\mathcal{R}_{\pi}$ operation is not a symmetry.  Then, the charge at the interface can take any value~\cite{ElectricPolarizNTGeometricPhase}.
Such a term may arise in CNTs if the $A$-$B$ symmetry is spontaneously broken due to many-body interactions~\cite{SCandCDWInstabilities,deshpande2009mott,Rontani2014}.

{Finally, we note that the $e/2$ interface charge is robust in the presence of disorder, as long as the bulk remains insulating and the symmetry under $\mathcal{R}_\pi$ is still maintained \emph{on average}.}



\begin{figure}[t]
	\centering
	\includegraphics[width=0.85\columnwidth]{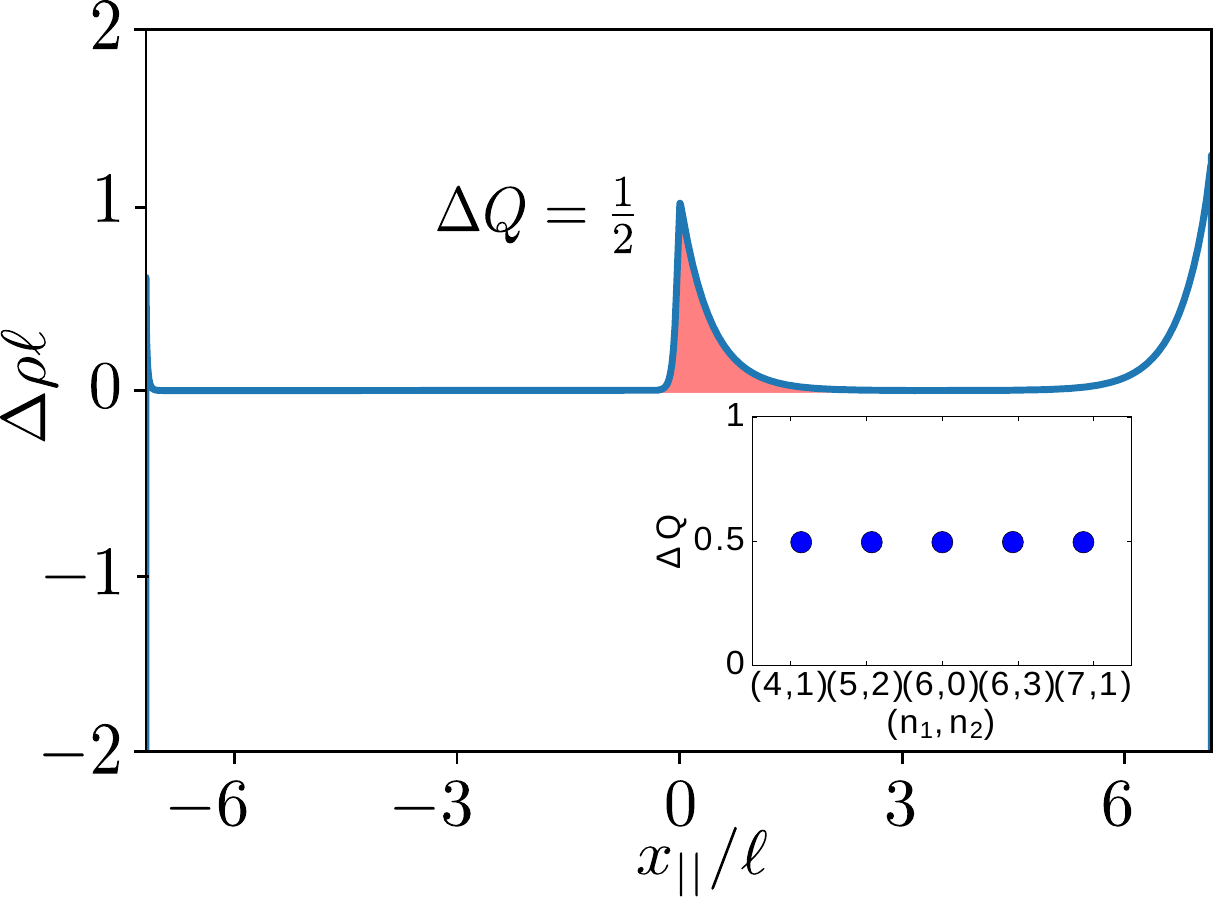}
	
	\caption{Charge density vs. position along the CNT, $x_\parallel$, for a $c=(6,0)$ (zigzag) CNT~\cite{SOM}. The tight-binding parameters used are $\tilde{t}=3$eV, $\tilde{\delta t}_{-}=14$meV, $\tilde{\delta t}_{+}=$84meV, $\tilde{\Delta}_{\mathrm{o}}=$4meV, $\tilde{\Delta}_{\mathrm{z}}=0.4$meV, $\tilde{\delta t}_{nnn}=$30meV. For these parameters, $\ell=\hbar v_F / \tilde{\Delta}_o\sim 0.2\mu$m. Inset: the interface charge, $\Delta Q$ for CNTs with different chiral vectors.}
	\label{fig:numerical_results}
\end{figure}

\emph{Numerical simulations.--}In order to demonstrate the phenomena described above, we simulated a tight-binding model of a CNT~\cite{SOM}. 
The model includes nearest-neighbor and next-nearest-neighbor hopping~\footnote{The next-nearest-neighbor hopping terms are not essential for the realization of a fractional charge. We have verified that the interface charge remains precisely quantized if the next-nearest neighbor hopping term is set to zero.}, as well as Zeeman and orbital-type spin-orbit coupling. 
The effect of strain is modelled by increasing the hopping on each bond by $\widetilde{\delta t} \cos^2(\alpha)$, where $\alpha$ is the angle between the direction of the bond and the chiral vector,  $\myvec{c}$. $\widetilde{\delta t}$ is taken to have the following spatial dependence:
$\widetilde{\delta} t(x_{||})= {\widetilde\delta t}_- + (\widetilde{\delta t}_+ - \widetilde{\delta t}_-)/ (1+e^{x_{||}/d})$, Where L is the CNT length. We used $d=10a$, where $a$ is the lattice spacing~\footnote{The simulations of chiral nanotubes were done using the Kwant package. See: C. W. Groth, M. Wimmer, A. R. Akhmerov, and X. Waintal, New J. Phys. \textbf{16}, 063065 (2014).}.

Fig.~\ref{fig:suggested_system}c shows the excess charge 
in the interface region, as a function of magnetic field and chemical potential, $\mu$.  The excess charge is summed over a region of length $\frac{L}{3}$ centered around the interface. For the given parameters this length corresponds to about $6\ell(B=B_0)$, where $\ell(B) = \hbar v_F / E_\mathrm{gap}(B)$ is the length scale associated with the gap $E_\mathrm{gap}(B)$ between the conduction and valence bands at $x_\parallel>0$. The field $B_0$ was tuned such that $B_{c\downarrow}<B_0<B_{c\uparrow}$.  
In this simulation, we set $\widetilde{\Delta}^{SO}_\mathrm{z}=0$ for clarity; it is included in the following.


In Fig.~\ref{fig:numerical_results} we show the excess charge density 
as a function of position, for a magnetic field $B_{c\downarrow}<B_0<B_{c\uparrow}$, in a $(6,0)$ zigzag CNT. Near the interface, there is a charge of $e/2$, localized over a region whose length is of the order of $\ell(B=B_0)$; another charge of $e/2$ is localized near the right end of the CNT. The inset shows the interface charge for CNTs with different chiral vectors; the charge is always $e/2$, independent of the chiral vector. 

\emph{Discussion.--}
We now discuss considerations for an experimental realization of our proposed setup. Refs.~\cite{ControllingEdgeStatesUsingABFlux, SO_CNT_meas_2,SO_CNT_meas_1,SO_CNT_meas_3} demonstrated that it is possible to close and reopen semiconducting gap in metallic CNTs using magnetic fields of a few Tesla. 
Realizing an interface charge of $e/2$ requires to drive \emph{only one} spin species through the topological transition. In this case, the maximum gap is determined by the strength of the spin-orbit coupling in the CNT, which has been estimated to be of the order of $\Delta^{SO}_{\mathrm{o,z}} \sim 0.08-1.7\mathrm{meV}$~\cite{SO_CNT_meas_2,jespersen2011gate,SO_CNT_meas_3,SO_CNT_meas_1}. Using $v_F\approx 10^6\, \mathrm{m/s}$, we get that the interface charge is localized in a region of length $\ell \sim \hbar v_F / \Delta^{SO}_\mathrm{o}\sim 0.35-7.5 \mu\mathrm{m}$.
Thus, 
our proposal requires a CNT whose length is a few microns. The possibility of fabricating long, pristine CNTs, and detection of localized charges, have been demonstrated in Refs.~\cite{waissman2013realization,hamo2016electron}.

To estimate the strain-induced gap in the CNT, we note that
the tension 
at the contact with the wedge is expected to be $\sim10\mathrm{nN}$~\cite{whittaker2006measurement}. Using the Young modulus of CNT, $0.1-1\mathrm{TPa}$~\cite{lau2006critical}, and the derivative of the energy gap with respect to strain, $\frac{100\mathrm{meV}}{0.01}$~\cite{minot2004determination}, the difference in the gap between the two regions is of the order of $\delta t_+-\delta t_-\sim1-10\mathrm{meV}$.

The existence of an interface charge of $e/2$ depends on the \emph{type} of spin-orbit coupling in the CNT.
 Keeping the system insulating in the range of magnetic field where only one spin species has an inverted mass requires that the orbital-type spin orbit coupling, $\Delta^{SO}_\mathrm{o}$, is larger in magnitude than the Zeeman type term, $\Delta^{SO}_{\mathrm{z}}$. Theoretical considerations suggest that $\Delta^{SO}_{\mathrm{z}}\sim\cos(3\theta)$, where $\theta$ is the chiral angle of the CNT~\footnote{$\theta$ is defined as the angle between the chiral vector of a CNT and a chiral vector corresponding to a zigzag CNT.}, while $\Delta^{SO}_{\mathrm{o}}$  does not dependent on $\theta$~\citep{SO_CNT_2}. In particular, $\Delta^{SO}_\mathrm{z}$ vanishes for an armchair CNT. Thus, the optimal chiral angle for realizing a charge of $e/2$ is close to $\theta = \pi/6$. This way, $\Delta^{SO}_\mathrm{z}$ is small, while the $K$ and $K'$ points are still crossed at different magnetic fields.

An important practical challenge in observing the soliton state in our proposed setup is the need to maintain the chemical potential in the gap throughout the CNT. This can be done by using an array of metallic gates~\cite{waissman2013realization}. If the wedge is made of a metal covered by an oxide insulting barrier, it can be used as an additional gate that can tune the JR state to the Fermi level~\cite{SOM}.

Finally, we comment on the effect of Coulomb interactions.
As long as the interactions are not strong enough to drive the system through a phase transition, the value of the interface charge is fixed to an integer multiple of $e/2$. The length scale $\ell$ over which the localized charge decays becomes longer in the presence of interactions. However, a rough estimate shows that this effect is small for realistic interaction strengths~\cite{SOM}.
In the presence of interactions, the induced charge density decays away from the wedge as $\rho_i\simeq\ell^{-1}(x_{||}/\ell)^{-3}$~\cite{SOM}, rather than exponentially.


\acknowledgements \emph{Acknowledgements.--}We thank Y. Baum, I. C. Fulga, Y. Oreg, and M. Rudner for useful discussions. EB acknowledges support from a Marie Curie CIG grant and from the European Research Council (ERC) under the European Union Horizon 2020 research and innovation programme (No. 639172). EB also thanks the hospitality of the Aspen Center for Physics, were part of this work was done. This work was supported in part by a grant from the Simons foundation. S.I. acknowledges financial support by the ERC Cog grant (See-1D-Qmatter, No. 647413).

\bibliography{Topological_transitions_CNT}

\end{document}


\title{Supplementary Information for: \\ Topological transitions and fractional charges induced by strain and magnetic field in carbon nanotubes}
\author{Yonathan Efroni}
\affiliation{Department of Condensed Matter Physics, Weizmann Institute of Science, Rehovot 7610001, Israel}
\author{Shahal Ilani}
\affiliation{Department of Condensed Matter Physics, Weizmann Institute of Science, Rehovot 7610001, Israel}
\author{Erez Berg}
\affiliation{Department of Condensed Matter Physics, Weizmann Institute of Science, Rehovot 7610001, Israel}
\affiliation{Department of Physics, James Frank Institute, University of Chicago, Chicago, Illinois 60637, USA}

\maketitle

\section{Tight binding Hamiltonian}\label{sec:tbHamiltonian}
\subsection{Setup of the model}
Here, we describe the tight-binding Hamiltonian used in our numerical simulations. {In the following, we denote the tight binding model parameters with an upper tilde symbol, to distinguish them from the corresponding parameters of the low energy effective Hamiltonian [Eq.~(1-2) in the main text].}
The Hamiltonian is written as
\begin{equation}
	H = H_0 + H_{Anis}+H_{SO}.
	\label{eq:tb}
\end{equation}
$H_0$ includes nearest-neighbour and next-nearest-neighbour hopping matrix elements:
\begin{equation}
	H_0 = -\sum_{ ij } t_{ij} c_i^{\dagger} c^{\vphantom{\dagger}}_j+ h.c.,
\end{equation}
where $t_{ij}=\widetilde{t} e^{iA_{ij}}$ for nearest neighbor sites, $t_{ij} = \widetilde{t}_{nnn} e^{i A_{ij}}$ for next-nearest neighbors, and $t_{ij}=0$ otherwise. $A_{ij} = \frac{2\pi \phi}{\phi_0} \frac{(\myvec{r}_i - \myvec{r}_j)\cdot\myvec{c}}{|c|^2}$ (where $\myvec{c}$ is the chiral vector, and $\myvec{r}_i$ is the position of site $i$ on the two-dimensional plane formed by opening the CNT and flattening it). This phase factor to a flux $\phi$ threaded through the CNT. $c^\dagger_i$ creates an electron on size $i$ of the lattice. 

$H_{Anis}$ describes the hopping anisotropy introduced by strain:
\begin{equation}
	H_{Anis} = -\sum_{\langle i,j \rangle} \delta t_{ij} c_{i}^{\dagger}c^{\vphantom{\dagger}}_{j} +h.c.
\end{equation}
The anisotropic part of the hopping, $\delta t_{ij}$, depends on the direction of the bond between the sites $i,j$ relative to the CNT axis. It is given by:
\begin{equation}
\delta t_{ij} = \widetilde{\delta t} \left(\frac{(\myvec{r}_i-\myvec{r}_j) \cdot {\myvec{c}}} {|\myvec{r}_i-\myvec{r}_j| |\myvec{c}| } \right)^2.
\label{eq:dt}
\end{equation}
Here, $\widetilde{\delta t}$ is proportional to the strain along the CNT axis. The anisotropy term was taken to be non-zero for nearest-neighbor sites only.

The Zeeman and orbital-type spin orbit coupling terms are included in $H_{SO}$, which is given by
\begin{equation}
	H_{SO} = \sum_{\langle i,j \rangle } i \Delta^{SO}_{\mathrm{o},ij} c_{i}^{\dagger} S_z c^{\vphantom{\dagger}}_{j} +
	\sum_{\langle \langle {i},{j} \rangle \rangle }i \Delta^{SO}_{\mathrm{z},ij} c_{i}^{\dagger} S_z c^{\vphantom{\dagger}}_{j}+ h.c.
\end{equation}
Here, $S_z$ is the spin component along the CNT axis. $\langle \cdot \rangle$ and $\langle \langle \cdot \rangle \rangle $ represent a summation over nearest- and second-nearest neighbor sites, respectively.
 The nearest-neighbor term leads to the $\Delta^{SO}_\mathrm{o}$ term in the low-energy theory [see Eq.~(2) of the main text], while the next-nearest neighbor term leads to the $\Delta^{SO}_{\mathrm{z}}$ term.
The amplitudes for these couplings depends on the bond direction as:
\begin{equation}
\Delta^{SO}_{\mathrm{z}/\mathrm{o},ij} = \widetilde{\Delta}^{SO}_{\mathrm{z}/\mathrm{o}} \left(\frac{(\myvec{r}_i-\myvec{r}_j) \cdot \myvec{c}}{|\myvec{r}_i-\myvec{r}_j|
|\myvec{c}|}\right)^2 \mathrm{sign}[(\myvec{r}_i-\myvec{r}_j) \cdot  \myvec{c}].
\end{equation}

{The tight-binding model parameters are related to those of the effective low-energy Hamiltonian [Eq.~(1) in the main text]. The relation can be derived by expanding the tight-binding Hamiltonian in momentum space around the $K$ and $K'$ points.
We quote the result below:}
\begin{align}
\delta t =& \frac{3}{4}\cos(3\theta) \tilde{\delta t}, \\
\Delta_{\mathrm{o}} =& \frac{1}{4}(3\sqrt{3}\cos(\theta)-4\sin(\theta)^3) \tilde{\Delta_{\mathrm{o}}}, \\
\Delta_{\mathrm{z}} =& \frac{\sqrt{3}}{2}\cos(3\theta) \tilde{\Delta_{\mathrm{z}}}.
\end{align}
\subsection{Relation to the SSH model}\label{sec:RelationToSSH}

We now specialize to the case of a zigzag CNT, and show that the Hamiltonian is directly related to the SSH model of a dimerized chain. The dimerization is determined by the hopping anisotropy $\delta t$ and by the magnetic field.
\begin{figure}[t]
	\centering
	\includegraphics[width=0.5\columnwidth]{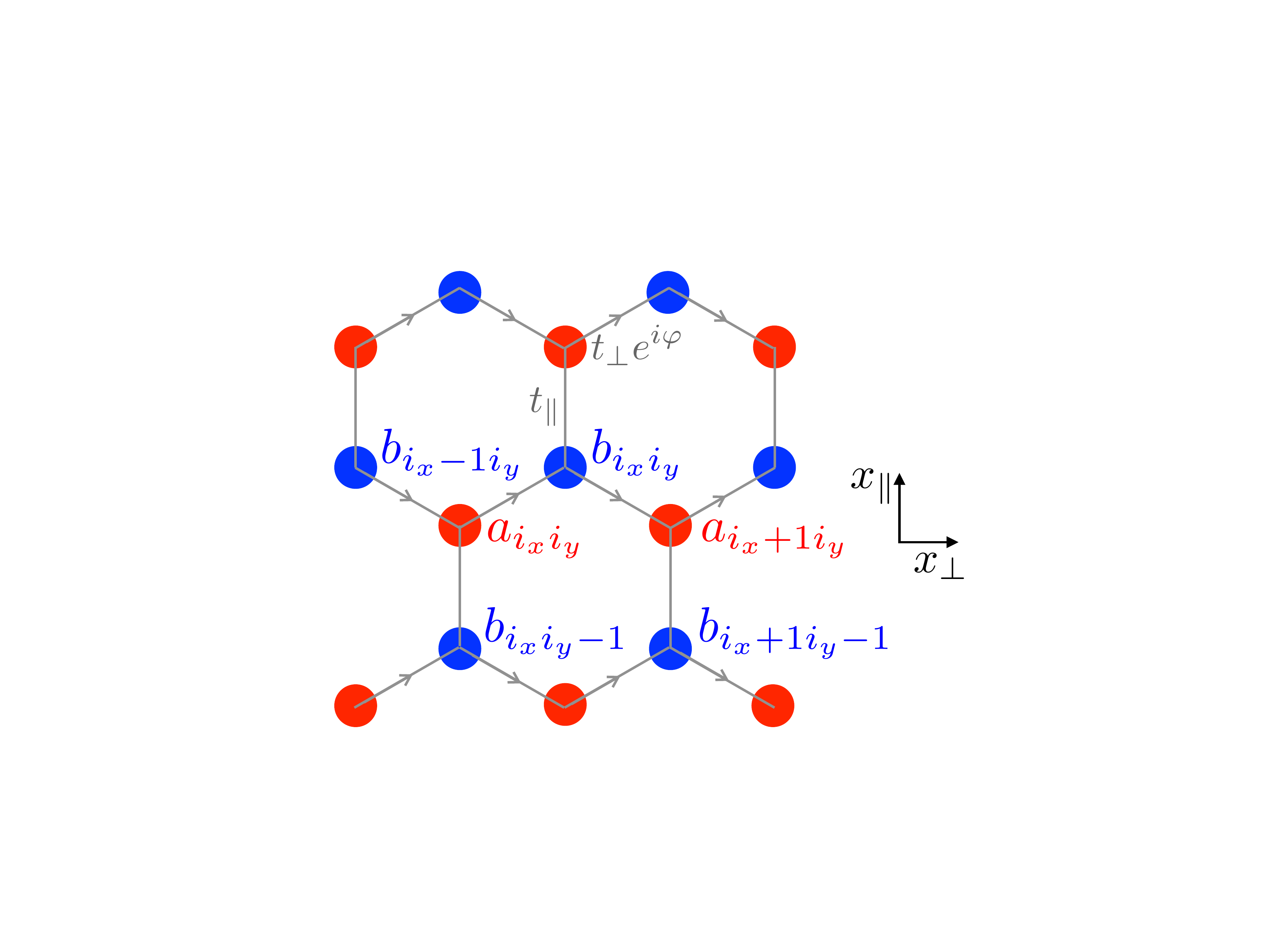}
	\centering
	\caption{Site labelling in a zigzag CNT. The $t_\perp$ hopping amplitudes have a phase $\varphi = \pi \phi/N\phi_0$, where $\phi$ is the flux through the CNT. $t_{||}$ and $t_\perp$ are defined in Eq.~(\ref{eq:5}). }
	\label{fig:zigzag}
\end{figure}
Let us consider a tight-binding model~(\ref{eq:tb}) of a zigzag CNT with a chiral vector $(N,0)$. For simplicity, we focus on one spin species only, and neglect the spin-orbit coupling and the next-nearest neighbor hopping. We label the sites of the zigzag CNT as shown in Fig.~\ref{fig:zigzag}, and denote the creation operators on the $A$, $B$ sublattices by $a_{i_x i_y}$ and $b_{i_x i_y}$, respectively, where $(i_x,i_y)$ label the unit cells. The Hamiltonian is of the form:
\begin{eqnarray}
H&=&-\sum_{ i_x,i_y } t_{\parallel} b_{i_x i_y}^{\dagger} a^{\vphantom{\dagger}}_{i_x i_y+1} + h.c. \\ 			
  &-&\sum_{ i_x,i_y } t_{\perp} e^{-i\varphi} (a_{i_x i_y}^{\dagger} b^{\vphantom{\dagger}}_{i_x i_y} + b_{i_x i_y}^{\dagger} a^{\vphantom{\dagger}}_{i_x+1 i_y})+ h.c. \nonumber \\
 &=&-\sum_{ k_\perp, i_y } \left( t_{\parallel} b_{k_\perp i_y}^{\dagger} a^{\vphantom{\dagger}}_{k_\perp i_y+1} + t_{k_\perp}(\phi) a^\dagger_{k_\perp i_y} \nonumber
 b^{\vphantom{\dagger}}_{k_\perp i_y} \right) + h.c.
\label{eq:5}
\end{eqnarray}
Here, $\varphi \equiv  \frac{\pi \phi }{N \phi_0}$.
In the third line, we have Fourier transformed perpendicular to the CNT axis, and introduced $t_{k_\perp}(\phi) \equiv 2t_\perp \cos\left(\frac{k_{\perp}(\phi)}{2}\right)$.  The tunneling matrix elements parallel and perpendicular to the CNT are given by $t_\parallel = t + \delta t$ and $t_\perp = t + \frac{\delta t }{4}$, respectively [see (Eq.~\ref{eq:dt})]. The perpendicular momentum is quantized according to $k_{\perp}(\phi)=\frac{2\pi}{N}(n+\frac{\phi}{\phi_0})$  where $n\in\{0,..N-1\}$.
The problem is equivalent to a set of $N$ decoupled 1D Hamiltonians, labelled by $k_\perp$, where each Hamiltonian describes a 1D dimerized chain, similar to the SSH-model.

If the hopping of the dimerized chain varies \emph{spatially}, such that $\mathrm{sign}[t_{\parallel}-t_{k_\perp}(\phi)]$ changes between two regions, a Jakiw-Rebbi soliton carrying a charge $\pm e/2$ emerges at the boundary between the regions. By applying a spatially varying strain, one can drive such a change. To see this, we assume that $N = 3m$ where $m$ is an integer (corresponding to a metallic CNT). Then, to leading order in $\delta t$ and $\phi$, the sign of the dimerization of the two sub-bands with transverse momentum closest to the Dirac points, corresponding to $\pm K_\perp = \pm \frac{2\pi m }{N}$, is given by
 \begin{equation}
 \label{eq:7}
t_{\parallel} -  t_{\pm K_{\perp}}(\phi)  \approx \frac{3}{4} \delta t \pm  \frac{\sqrt{3}\pi \phi }{N \phi_0} t.
 \end{equation}
Thus, changing $\delta t$ spatially can change the sign of the dimerization of one of the sub-band at $k_\perp = \pm K_\perp$, driving it through a topological transition, while all the other sub-bands remain gapped.

\begin{figure}[t]
	\centering
	\includegraphics[scale=0.5]{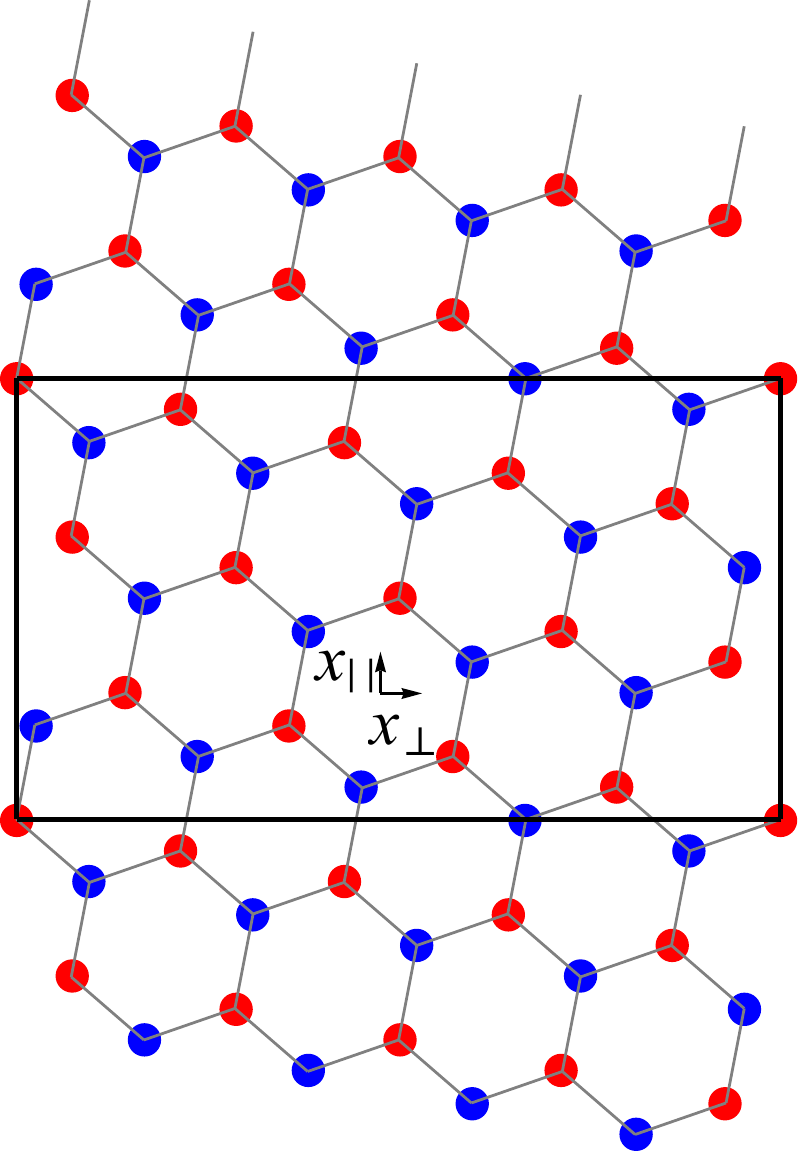}
	\centering
	\caption{$\myvec{c} = (4,1)$ chiral CNT. The black rectangle marks the unit cell. The CNT is invariant under the $\mathcal{R}_{\pi}$ operation, which amounts to a $\pi$ rotation around the center of a hexagon. This operation interchanges the $A$ and $B$ sub-lattices (red and blue disks, respectively). }
	\label{fig:RpiSymmetry}
\end{figure}

\section{$\mathcal{R}_{\pi}$ symmetry of CNT}\label{sec:RpiSymmetry}
Any CNT (both chiral or a-chiral) is symmetric under a $180^{\circ}$ rotation around a perpendicular direction to the CNT axis that passes through a center of a hexagon or through the center of a bond. This can be most easily seen by noting that the inversion operation for the two-dimensional honeycomb lattice is equivalent to a $180^\circ$ rotation in the case of a CNT. In the honeycomb lattice, inversion applies the transformation $x\rightarrow -x,y\rightarrow -y$, while for a CNT,  a $180^\circ$ rotation applies the transformation  $x_{\parallel}\rightarrow -x_{\parallel},x_{\perp}\rightarrow -x_{\perp}$ (see Fig.~\ref{fig:RpiSymmetry}). Both transformations interchange the $A$ and $B$ sublattices.

In the presence of an axial magnetic field through the CNT, a $180^\circ$ rotation also changes $\myvec{B} \rightarrow -\myvec{B}$. The system is symmetric under a $180^\circ$ rotation followed by time reversal. This operation is denoted by $\mathcal{R}_\pi$ in the main text.



\section{Fractional charge at the edges}\label{sec:UniformStrain}

We have shown that in our system, a domain wall between spatial regions with different strain can carry a Jackiw-Rebbi soliton with a fractional charge. It is natural to ask whether a another fractional charge can also appear \emph{at the edge}. One would natively expect this to be the case, since overall, the total charge of an isolated CNT must be an integer multiple of $e$. Moreover, one can consider a setup in which the strain of the CNT is spatially uniform, and ask whether a well-defined fractional charge can appear at the edges.

The existence of edge charges is subtle, however, since it may depend on details of the edge. Consider, for example, a dimerized SSH chain. In this case, an edge state carrying a charge of $e/2$ \emph{may or may not} appear at the edge, depending on how the chain is terminated. One can eliminate the edge state, as well as its associated charge, by adding a single site to the chain.
This may seem surprising, since the charge at the edge is related to the \emph{bulk} polarization; however, the latter quantity quantity depends on the definition of the unit cell. In the SSH chain, the bulk polarization can either be $0$ or $e/2$ (and, correspondingly, the Zak phase can be either 0 or $\pi$), depending on how the unit cell is defined. Thus, the appearance of a fractional charge at an edge of a spinless SSH chain, and by analogy in our proposed system, is in general not a robust bulk property.

However, if the magnetic field is tuned such that a single spin species crosses the topological transition, while the other remains in the trivial phase (which is possible thanks to the presence of spin-orbit coupling), then the $e/2$ edge charge is robust, and is \emph{completely} independent of the structure of the edge. In order to understand this, let us consider the case of two SSH chains (which represents the two spin flavors). If the two SSH chains are in different topological phases (i.e., the dimerization of the hopping has an opposite sign in the two chains), then at the edge, one of the two SSH chains necessarily has an edge state carrying an $e/2$ charge, while the charge at the end of the other chain must be an integer times $e$. This is independent of the particular edge termination. The appearance of an $e/2$ charge follows from the fact that the bulk polarization is $e/2$, independently of how the unit cell is defined.

In contrast, in the situation where both spin flavors have the same polarization, a spin-charge separated state may either appear at the edge or not, depending on the way the system is terminated. At a domain wall between the two states with an opposite dimerization, a spin-charge separated state appears independently of details.

\section{Effect of Coulomb interactions}\label{sec:UniformStrain}

In the main text, we have shown that the existence of the $e/2$ charge is protected by the $\mathcal{R}_{\pi}$ symmetry. As long as the Coulomb interaction does not drive the system through a phase transition, the $\mathcal{R}_{\pi}$ symmetry is still present. Thus, the value of the interface charge is fixed to an integer multiple of $e/2$ even in the presence of interactions. However, as we now discuss, the charge distribution around the interface depends on the interactions.

\begin{figure}[t]
	\centering
	\includegraphics[scale=0.32]{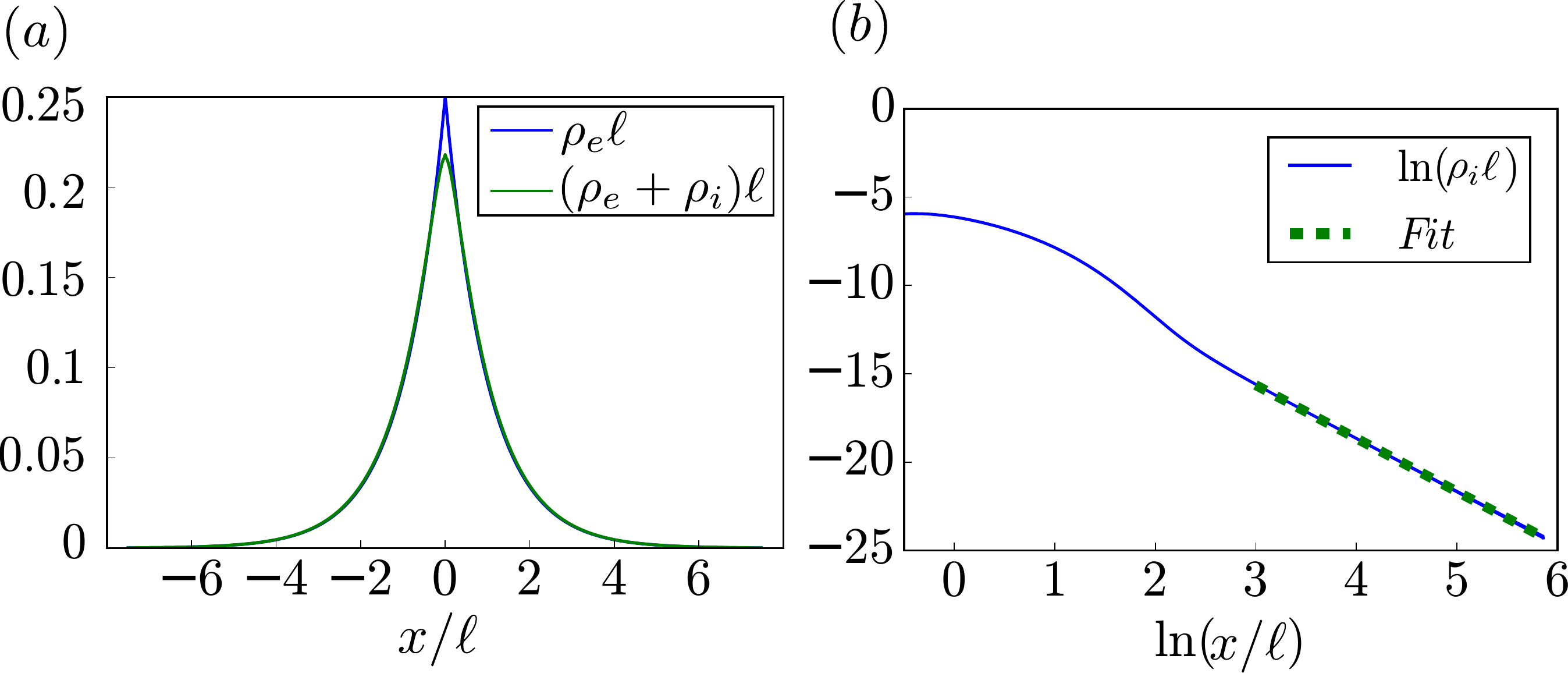}
	\centering
	\caption{(a) The total charge($\rho_{i}+\rho_{e}$) and external charge($\rho_{e}$), versus real space distance scaled by $\ell$,      $z/\ell$. The calculation was done by using $\alpha=e^2/\hbar v_F=3$, $r_c=0.001$,$\ell=1$.(b) Log-Log plot of the induced charge versus real space distance scaled by $\ell$, $z/\ell$, and linear fit to the tail of the function. The linear fit is of the form $\ln\left(\rho_{i}\right)=-a_1\ln\left(z\right)+a_0$. The values extracted from the fit are $a_1=3.006,a_0=-7.73$. The decay character decays with a power law of 3.}
	\label{fig:interaction}
\end{figure}

We consider a massive 1D Dirac Hamiltonian:
\begin{equation}
H=\int dk\, \left( \hbar v_F k \sigma_x+m \sigma_y \right).
\end{equation}
The charge susceptibility as a function of the wavevector $q$ can be expressed as: 
\begin{equation}
\chi(q,\omega) = -\int \frac{dk}{2\pi} \frac{d\omega}{2\pi} \mathrm{Tr}\left[ G(k,\omega) G(k+q,\omega) \right],
\end{equation}
where the Matsubara Green's function is given by $G(k,\omega) = (i\omega - \hbar v_F \sigma_x -m \sigma_y)^{-1}$. The $\omega$ integral can be performed to give
\begin{equation}
\chi(q)=\frac{\ell}{4\pi\hbar v_F}\int dk\ \frac{\Large|1-e^{i \left(\tan^{-1}\frac{1}{\ell k+\ell q}-\tan^{-1}\frac{1}{\ell k} \right)} \Large|^2}{\sqrt{(\ell k)^2+1}+\sqrt{(\ell k+\ell q)^2+1}},
\label{fig:interaction_res}
\end{equation}
 where $\ell=\hbar v_F/m$. 
We add an external test charge representing the interface $e/2$ charge between the two topological phases.
In the absence of interactions, the distribution of the interface charge is given by $\rho_e(x)= \frac{1}{4\ell} e^{-|x|/\ell}$.
We now compute the correction to this distribution in the presence of Coulomb interactions. The effective Coulomb potential as a function of separation, $v(x)$,  is the electrostatic interaction between two rings of charge. We model it by the function:
\begin{equation}
 v(x)= e^2\left[\frac{1}{\sqrt{x^2+r_c^2}} - \frac{e^{-|x|/r_c}}{r_c}\log(|x|/r_c)\right],
\end{equation}
 where $r_c$ is of the order of the CNT radius (i.e., $r_c\ll \ell$). At small distances, $v(x\ll r_c) \sim -\log(|x|/r_c))$, while at large distances $v(x \gg r_c)\sim  1/x$, which is the expected asymptotics of the Coulomb interaction between rings.

Treating the problem within the random phase approximation (RPA), and assuming that linear response holds, the Fourier transform of the induced charge, $\rho_i(q)$, satisfies:

\begin{equation}
\rho_i(q)=-\frac{v(q)\chi(q)}{1+v(q)\chi(q)}\rho_e(q)
\label{eq:rho_i_q}
\end{equation}

By taking the inverse Fourier transform we can evaluate the form of the total charge $\rho(x)=\mathcal{F}^{-1}[\rho_e(q)+\rho_e(q)](x)$. The total charge as a function of position, calculated numerically using Eqs.~(\ref{fig:interaction_res},\ref{eq:rho_i_q}) with realistic parameters, is shown in Fig.~\ref{fig:interaction}a. We find the Coulomb interactions make the charge distribution broader, due to the induced screening charges in the CNT. However, as seen in the figure, this effect is found to be quantitatively small.

The long-distance behavior of the distribution is qualitatively modified in the presence of interactions. We can extract this behavior by examining Eq.~(\ref{eq:rho_i_q}) at small $q$. We find that $\rho_i(q\rightarrow 0)\propto-\alpha (\ell q)^2 \log(|q|)\rho_e(0)$, with $\alpha=e^2/\hbar v_F $. This leads to $\rho_i(x \rightarrow \infty)\propto \alpha \ell^{-1} (x/\ell)^{-3}$. This behavior is verified in Fig.~\ref{fig:interaction}b.

\section{Maintaining the chemical potential in the gap of the CNT}\label{sec:maintain}

A major challenge in realizing the proposed setup experimentally is the need to maintain the chemical potential within the topological gap throughout the CNT. To achieve that, the chemical potential needs to be controlled with high precision, both around the wedge and in the bulk of the CNT. Here, we briefly discuss a possible design of an experimental setup that can achieve this. 

The potential landscape can be controlled by using an array of metallic gates underneath the nanotube. A similar setup has been used, for example, in Ref.~\cite{waissman2013realization}. Controlling the voltage of each gate individually allows to ``flatten'' the potential along the nanotube.

\begin{figure}[h]
\centering{
\includegraphics[width=0.45\textwidth]{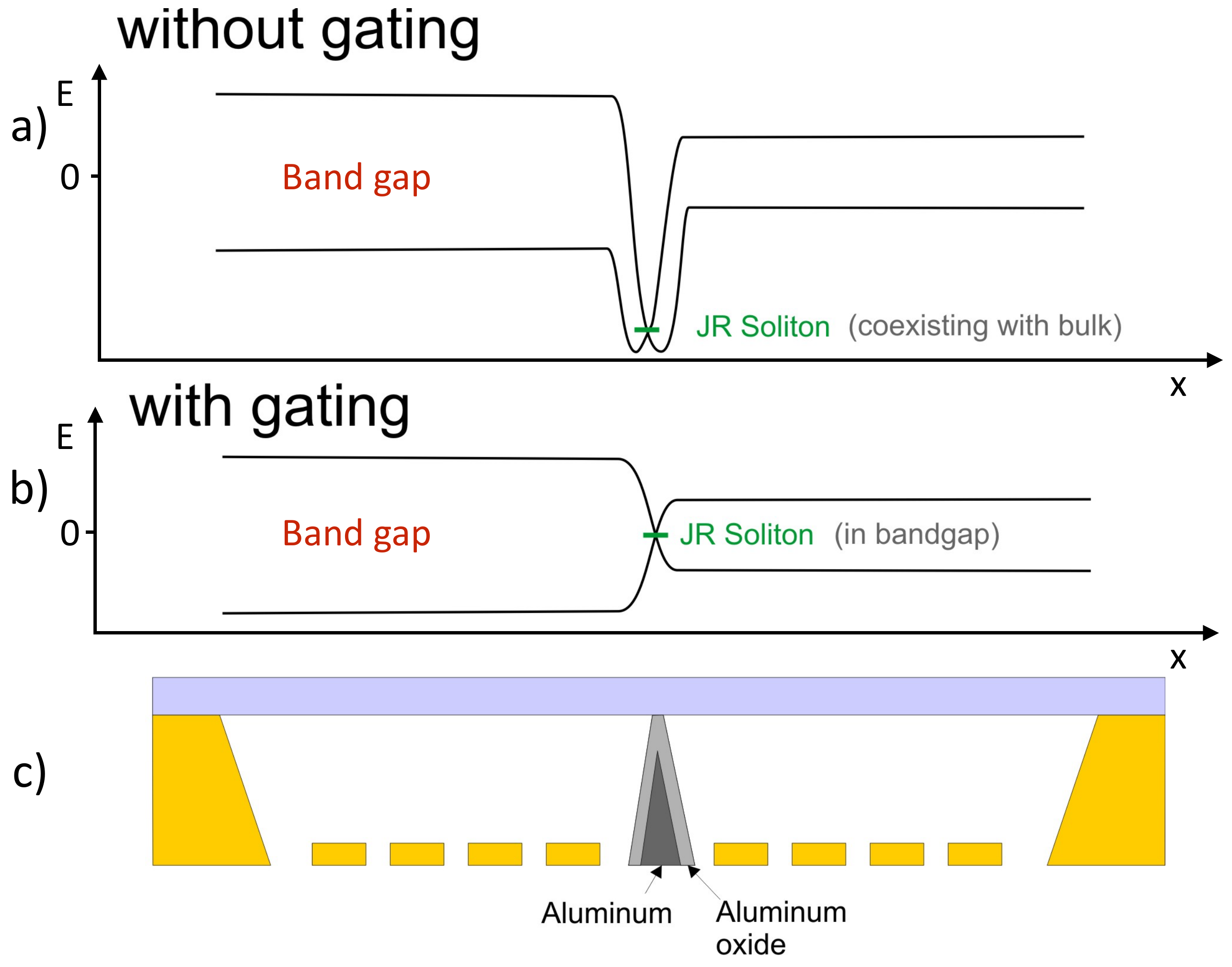}
\caption{a) Energy diagram for the two bands of the nanotube as a function of position, in a setup where the potential of the wedge is uncontrolled. In the bulk of the nanotube, the chemical potential is within the gap. However, the wedge distorts the potential around it, pushing the Jackiw-Rebbi soliton down in energy into the valence band, where it can mix with bulk states.
b) If the wedge is metallic (but electrically insulated from the nanotube by a thin oxide layer), it can be voltage biased with respect to the nanotube, bringing the JR soliton state back into the energy gap.
c) Schematic design of the proposed system. In addition to the metallic gates used to flatten the potential landscape in the nanotube, the wedge (made of aluminum covered by a thin oxide layer) is used as an additional gate, bringing the JR state near the middle of the gap.}
\label{fig:topView}}
\end{figure}

The wedge at the middle of the nanotube is expected to act as an electrostatic perturbation, distorting the electric field lines around it. This is due to the fact that the wedge's dielectric constant is different from that of its surrounding. Moreover, the work function of the wedge is different from that of the nanotube, bending the bands of the nanotube near it and causing charge to transfer from the wedge to the nanotube.
As a result, the potential near the wedge can be very different from the that of the bulk of the nanotube, and the Jackiw-Rebbi state bound to the soliton could sink below the gap or rise above it - see Fig.~\ref{fig:topView}a. In that case, the JR bound state mixes with the bulk states at the same energy, making it difficult to observe. [Note that the charge bound to the wedge region is still $(n+1/2)e$, regardless of the potential near the wedge, as long as the chemical potential is within the gap in a sufficiently large region away from the wedge.]

In order to be able to tune the Jackiw-Rebbi state back to the Fermi level and to tune the interface charge to precisely $e/2$, the wedge itself can be made from a metal that can be voltage biased relative to the nanotube, covered by a thin oxide that insulates it from the nanotube.
Such a setup is illustrated in Fig.~\ref{fig:topView}c: the wedge is made of aluminum, covered by a thin layer of aluminum oxide. As explained above, an array of metallic gates is used to flatten the potential along the nanotube, while the voltage of the wedge is tuned such that the Jackiw-Rebbi state is near the middle of the bulk band gap (as illustrated in Fig.~\ref{fig:topView}b). This setup allows for enough control to pin the Fermi energy to the gap throughout the device.

\bibliography{Topological_transitions_CNT}